\documentclass[letterpaper,twocolumn,prl, showpacs]{revtex4}
\usepackage{graphicx, amsmath,amssymb, multirow}

\newcommand{\be}[0]{\begin{equation}}
\newcommand{\ee}[0]{\end{equation}}

\begin{document}

\title{Quantum spatial superresolution by optical centroid measurements}
\author{Heedeuk Shin$^1$, Kam Wai Clifford Chan$^{1,2}$, Hye Jeong Chang$^{1,3}$, and Robert W.
Boyd$^{1,4}$}
\affiliation{$^{1}$The Institute of Optics,
University of Rochester, Rochester, NY 14627, USA}
\affiliation{$^{2}$Rochester Optical Manufacturing Company,
Rochester, NY 14606, USA}
\affiliation{$^{3}$Korean Intellectual
Property Office, Daejeon 302-701, Korea}
\affiliation{$^{4}$Department of Physics, University of Ottawa,
Ottawa, ON K1N 6N5, Canada} \email{boyd@optics.rochester.edu}
\date{\today}

\begin{abstract}
Quantum lithography (QL) has been suggested as a means of achieving enhanced spatial
resolution for optical imaging, but its realization has been held back by the low
multi-photon detection rates of recording materials. Recently, an optical centroid
measurement (OCM) procedure was proposed as a way to obtain spatial resolution enhancement
identical to that of QL but with higher detection efficiency (M. Tsang, Phys. Rev. Lett.
102, 253601, 2009). Here we describe a variation of the OCM method with still higher
detection efficiency based on the use of photon-number-resolving detection. We also report
laboratory results for two-photon interference. We compare these results with those of the
standard  QL method based on multi-photon detection and show that the new method leads to
superresolution but with higher detection efficiency.
\end{abstract}

\pacs{}

\maketitle

The spatial resolution of optical imaging systems has traditionally been considered to be
limited by the Rayleigh resolution criterion.  One means of overcoming this limit
\cite{fonseca} is to make use of the photon correlations that exist in certain quantum
states of light.  A specific example of such an approach is the quantum lithography (QL)
proposal of Dowling and coworkers \cite{boto}. In this approach, a path-entangled state of
$N$ photons (a $N00N$ state) is used to write an interference pattern onto a recording
material that responds by means of multi-photon absorption (MPA), producing
$N$-fold-enhanced resolution as compared with a classical fringe pattern. Experimental
procedures for creating $N00N$ states with up to $N=5$ photons by means of spontaneous
parametric down conversion have been reported by several group
\cite{hom,steinberg,takeuchi,walther,afek}. Experimental demonstrations of spatial
superresolution through the QL procedure have, however, been rather limited.   In one
approach, multi-photon absorbing (MPA) lithographic materials are mimicked by using two
single-photon detectors operated in coincidence \cite{shih,takeuchi2}.  In another,
poly(methyl-methacrylate) was used as a MPA lithographic material for recording
sub-Rayleigh interference patterns, but only when excited by intense classical light
\cite{chang}.  In order to realize true quantum lithography, very sensitive lithographic
materials that can respond by MPA to weak quantum states of light are required. The use of
time-energy-entangled multi-photons can provide significant enhancement of the MPA
transition rate because of the near-zero variation in birth time \cite{gea}.  There has
also been some uncertainty in the trade-off between resolution enhancement and MPA
enhancement \cite{steuernagel,tsang1,tsang2,plick}.   In summary, true quantum lithography
has yet to be realized because of the low MPA efficiency of available materials.

Recently, Tsang  \cite{tsang3} proposed an  ``optical centroid"
method for achieving spatial interferometric superresolution with
much higher detection efficiency than that of QL. Instead of using
detectors that respond by MPA as in quantum lithography, an array
of single-photon detectors followed by postprocessing is used. The
addresses of the $N$ detectors that fire in response to $N$
incident photons are recorded,  and the centroid of the positions
of those detectors is computed. A histogram of the positions of
optical centroids determined by repeated measurements is then
produced. This histogram shows an interference pattern with a
resolution enhancement identical to that of the QL method. If the
pixel size of the detector array is much smaller than the
correlation area of the entangled photons, the probability that
the photons arrive at different pixels is much larger than the
probability that they arrive at same pixel. For this reason, the
detection efficiency of the OCM method is much higher than that of
the QL method, which relies on MPA.

In this Letter, we report the results of a proof-of-principle
experiment that demonstrates optical superresolution based on an
improved version of Tsang's OCM method.  The improvement comes
about by implementing a form of photon-number-resolving (PNR)
detection, which leads to still higher efficiency than Tsang's
original proposal.   The interference fringes obtained by the OCM
method are found to show resolution enhancement identical to that
of the QL method, but with higher detection efficiency. To the
best of our knowledge, ours is the first experimental
demonstration of spatial resolution enhancement using the OCM
method.

We next briefly review the theory of resolution enhancement by
both the OCM  and QL methods. Under the one-dimensional ($x$) and
monochromatic approximations \cite{tsang1}, the electric field
operator on the detection plane is given by
\begin{equation}
    \hat{E}^{(+)}(x) = i \sqrt{\frac{\eta}{(2\pi)^2}} \int d q \ \hat{a}(q)
    e^{i q x} ,
\end{equation}
where $\eta = \hbar/(2 \epsilon_0 c^2 T)$ with $T$ being the
normalization time scale.  Here $q$ and $x$ are respectively the
transverse wavevector and transverse position on the detector
plane. The $N00N$ state on the detector plane is given by
\begin{eqnarray}
    |N00N\rangle
    &=&
    \frac{1}{\sqrt{2 N!}}
    \left\{ \left[\hat{A}^\dagger\right]^N +
    \left[\hat{B}^\dagger\right]^N \right\}|0\rangle
\end{eqnarray}
where $\hat{A}^\dagger$ and $\hat{B}^\dagger$ represent the
annihilation operators of photons in modes $A$ and $B$ falling
onto the detector plane, respectively. $\hat{A}^\dagger$ is
$(1/\sqrt{\Delta\kappa}) \int d\kappa \ F^*\left[(\kappa_{A} +
\kappa) / \Delta\kappa\right] \hat{a}^\dagger(\kappa)$ with the
mean $x$-component of wavevector  $\kappa_A=\kappa_0$.
Analogously, $\hat{B}^\dagger$ has $\kappa_B=-\kappa_0$ and
$\triangle \kappa$ is the uncertainty of the transverse
wavevector. $F(q)$ is the normalized transverse wavevector profile
of the photon packet. Note that $[\hat{A}, \hat{A}^\dagger] =
[\hat{B}, \hat{B}^\dagger] = 1$ and $[\hat{A}, \hat{B}^\dagger] =
0$ for $\kappa_0 \gg \triangle \kappa$. If we set $F(q) =
(1/\sqrt{\pi}) \exp\left(-q^2/2\Delta \kappa^2\right)$,  the
$N$-photon, conditional probability density for the QL method
becomes
\begin{eqnarray}
     P_C(x)
&\!\!\!=&\!\!\!     \left\langle :
    \hat{I}(x)^N
    : \right\rangle
    =
\left\langle \left[\hat{E}^{(-)}(x)\right]^N
    \left[\hat{E}^{(+)}(x)\right]^N \right\rangle
\nonumber\\
    &\!\!\!\!\!\!\!\!\!\!\!\!\!\!\!\!\!\!\!\!\!\!\!\!=&\!\!\!\!\!\!\!\!
    \frac{N! \eta^N \Delta\kappa^N}{\pi^N}
    e^{\left(-N\Delta\kappa^2x^2\right)}
    \left[
    1+\cos\left(2N \kappa_0\, x\right)
    \right] .
\label{P_C}
\end{eqnarray}

  On the other
hand, the probability distribution for the optical centroid is
\begin{eqnarray}
\!\!\!\!\!\!P_M(X) &\!\!\!=&\!\!\! \int d \xi_1 \cdots d \xi_{N-1}
\left\langle : \prod_{n=1}^N \hat{I}(X+\xi_n) : \right\rangle
\nonumber\\
&\!\!\!\!\!\!\!\!\!\!\!\!\!\!\!\!\!\!\!\!\!\!\!\!=&\!\!\!\!\!\!\!\!\!\!\!\!
\frac{N! \eta^N \Delta\kappa}{\sqrt{N \pi^{N+1}}}
e^{\left(-N\Delta\kappa^2 X^2\right)} \left[ 1 + \cos\left(2N
\kappa_0\, X\right) \right] , \label{P_m}
\end{eqnarray}
where the centroid and relative-position coordinates are defined
as $X = \frac{1}{N}\sum_{n=1}^N x_n$ and $ \xi_n =  x_n- X$ with
$n=1, \cdots, N$, respectively. $P_M(X)$ is a marginal probability
density.  We see that they both give the same spatial resolution
enhancement. However, the ratio of
probabilities for the  two cases is
\begin{equation}
\frac{P_M(x) \delta x}{P_C(x) \delta x^N} = \frac{1}{\sqrt{N}}
\left( \frac{\sqrt{\pi}}{\Delta\kappa \delta x} \right)^{N-1},
\end{equation}
where $\delta x$ is the pixel size of detector. For typical situations in which the beam
size is much larger than the pixel size, one has $\Delta\kappa \delta x \ll 1$. Therefore,
$P_M(X)$ can be much greater than $P_C(x)$. Note that this conclusion holds even before
taking account of the greatly different detection efficiencies of single-photon and
multi-photon detectors.  When these differences are taken into account, the OCM method
becomes even more favorable.

To provide an intuitive understanding of the tradeoffs between the
QL and OCM methods, we next present an analysis based on the use
of combinatorics. We suppose that the correlation area of the
photon field is $M$ times larger than the pixel size on the
detector array and that $N$ entangled photons arrive at random
positions on the detector array within this correlation area. The
total number of combinations with repetition  for $N$ photons
falling on  $M$ pixels is $C_{\rm total} = (M+N-1)! / (N!(M-1)!)$.
Every such case occurs with equal probability $1/C_{\rm total}$
because of the assumption of random positions. Therefore, the more
combinations a particular method has, the more efficient it is.
For instance, quantum lithography requires $N$-photon absorption,
and the number of cases of $N$ photons falling onto the same pixel
is $C_{N\rm PA} = M$. If all $N$ photons do not fall onto the same
detector this event will be lost, leading to decreased detection
efficiency.  In the OCM method, however, a single-photon detector
array is used, and the number of combinations of detecting $N$
photons by $N$ different pixels among the $M$ pixels is $C_{\rm
SPA}=M!/(N!(M-N)!)$. For small pixel size or large correlation
area ($M \gg N$), the OCM method will be much more efficient than
the QL method ($C_{\rm total} \sim C_{\rm SPA} \gg C_{N\rm PA}$),
as Tsang predicted.

In the laboratory, however, practical concerns may preclude the
pixel area from being much smaller than the correlation area.
Moreover, most currently available  high-sensitivity detectors are
not photon-number resolving (PNR), that is, they cannot
distinguish between one and several photons falling onto the
detector. If more than one photon arrives at a given pixel, the
single-photon detector will count this as a single event, and
fewer  than $N$ detectors will register. Then the OCM protocol
will discard this event, leading to decreased measurement
efficiency.  This loss of efficiency becomes increasingly more
significant for large photon numbers $N$ or small values of $M$.

Loss of efficiency due to multi-arrivals at one pixel can be
eliminated by using a photon-number-resolving (PNR) detector
array. Recently, PNR detectors based on superconductive nanowire
technology with high quantum efficiency  have been developed
\cite{goltsman}. The PNR detector array will measure the addresses
of pixels that fire as well as the number of photons at these
pixels.   An accurate optical centroid of the detection process
can thereby be computed. The OCM method with a PNR detector array
has the number of combinations with repetition for $N$ photons
falling onto the array given by  $C_{\rm PNR}
=(M+N-1)!/(N!(M-1)!)$. This result indicates that the PNR detector
array can use all of the cases of $N$ photons arriving at the
detector. The PNR detector array will work like a MPA detector
array for $M \sim 1$ and will be almost equal to the single-photon
OCM detector for $M \gg N$.

We have performed experimental studies of superresolution for
two-photon interference (that is, $N=2$) for both the  QL and OCM
methods. The experimental setups  are the same for both cases
except for the detection method, as shown in Fig.~1.  A UV light
beam at 400-nm wavelength is generated by second-harmonic
generation of 100-fs pulses at 800-nm wavelength at repetition
rate of 82 MHz and is split into two beams by a beam splitter
(BS1).  A 1.5-mm-thick BBO crystal is placed in each (mutually
coherent)  UV beam, and spontaneous parametric down conversion
occurs randomly in each crystal under type 1 collinear phase
matching conditions. After blocking the pump beams using
interference filters (IF), the photon number state $|\Psi\rangle$
in mode $A$ and $B$  is given by
\begin{eqnarray}
|\Psi\rangle= |0\rangle_A|0\rangle_B + \gamma (|2\rangle_A
|0\rangle_B +|0\rangle_A|2\rangle_B)/\sqrt{2}+\ldots,
\end{eqnarray}
where $\gamma$ is the probability of creating a photon pair by
parametric down-conversion. At low pump power, $\gamma$ is much
smaller than unity,  and we can thus ignore multi-pair generation
proportional to higher powers of $\gamma$.  The generated light is
then well approximated as a two-photon $N00N$ state. This
procedure for generating the 2002 state is convenient because it
avoids the need for a maintaining the sensitive alignment of a
Hong-Ou-Mandel setup \cite{hom}.  Moreover,  strong photon flux
can be obtained by using long crystals or periodically poled
crystals \cite{dayan}.  We use a second beam splitter (BS2) to
combine these two beams with a small angle ($\theta\sim
0.033^\circ$) between them.  To increase the  collection
efficiency, two spherical lenses with 10-cm focal length are
located after each IF, and each spherical lens is defocused by 0.5
mm to make the correlation area larger than the pixel size. The
measured correlation area has a diameter of approximately 0.5 mm.

\begin{figure}
\centering
\includegraphics[width=8cm]{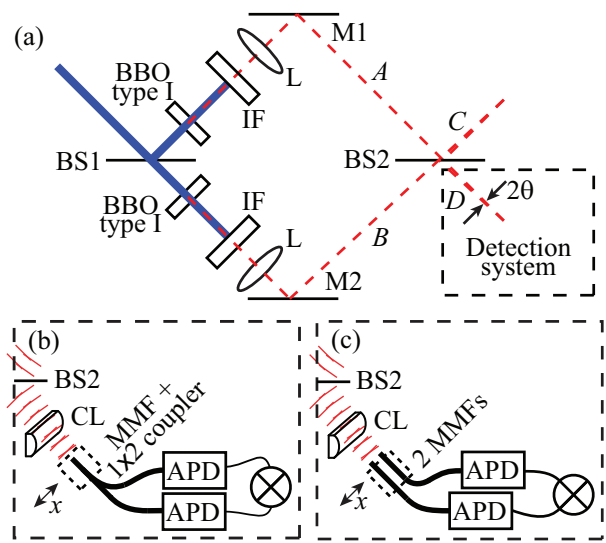}
\caption{(Color online) (a) Experimental setup for producing
two-photon interference. The dashed lines indicate the
down-converted photon-pair fluxes, and the dashed box represents
the detection system used to measure the two-photon interference
pattern.  (b) The detection system for the QL process.    A
coincidence measurement that mimics two-photon absorption.  (c)
Detection system for the OCM procedure, as described in the text.
In each case, a cylindrical lens (CL) is positioned in front of
the detection systems to increase collection efficiency and the
coincidence window time was 7 ns.  APD = avalanche photodiode. }
\end{figure}

The two detection systems were prepared using multimode fibers
(MMFs) with core and cladding diameters of 62.5 $\mu$m and 125
$\mu$m, respectively.  For the QL case (see Fig.~1 (b)), a single
MMF acting as a collector was scanned in discrete steps of 50
$\mu$m  across the detection region using a motorized translation
stage.  The output of this fiber was split into two additional
MMFs whose outputs were monitored by two single-photon detectors
operating in coincidence \cite{takeuchi2}.  The coincidence
circuit counts how many photon pairs arrive simultaneously at the
position of the input fiber, thus emulating a two-photon detector.

The OCM detection system was constructed as follows. According to
the OCM proposal, the detection system should consist of a linear
array of detectors each of which can respond with high sensitivity
to individual incident photons.  Detector arrays of this sort are
not readily available.  Instead, we simulated such a detection
system by using two MMFs of variable separation serving as
detectors, as shown in Fig.~1 (c).  For a given fixed separation,
this fiber pair is scanned through the detection region while
coincidence counts are recorded.  The coincidence circuit counts
how many photons arrive simultaneously at the two spatially
separated inputs.   The centroid position is located at the mean
position of the two fibers. This procedure is then repeated
sequentially for other fiber-to-fiber separations  of 125, 250,
375, 500, and 625 $\mu$m.  The coincidence count rates at a given
centroid position are then summed for all fiber separations.  In
this manner we collect the same data that would have been
collected (although more rapidly) by a multi-element detector
array.  To simulate a PNR detector array, we include the case of a
single collection fiber coupled to two single-photon detectors
(Fig. 1(b)) with that of two collection fibers of variable
separation (Fig. 1(c)).

Our experimental results are shown in Fig.~2.  In part (a) of the
figure, we show the form of the classical, single-photon
interference fringes.  These results were obtained using strongly
attenuated laser light of 800-nm wavelength, and serve as a
reference.  Under our experimental conditions, the period of these
classical interference fringes was 0.69 mm. Next, two-photon
interference fringes were recorded using the QL detector of
Fig.~1(b).  Both singles counts and coincidence counts are shown
in Fig.~2(b).  The singles counts show a Gaussian profile, whereas
the coincidence counts exhibit an interference pattern with a
decreased period of about 0.34 mm. Therefore, the QL method shows
a factor-of-two increase in spatial resolution as predicted
\cite{boto} and observed previously in references \cite{shih}  and
\cite{takeuchi2}.

Next, we repeated the measurement of the two-photon spatial
interference pattern using our OCM detection system.   The single-
and two-photon count rates for  MMFs  separated by 125 $\mu$m are
shown in Fig.~2(c). Because of the fiber separation, the
single-photon data have different peak positions separated from
each other by approximately 125 $\mu$m.  The period of the
two-photon fringes is approximately 0.34 mm, the same as the QL
result, demonstrating enhanced resolution by the OCM method. The
fitted curve for the two-photon coincidence counts is a sinusoidal
pattern weighted by a Gaussian function. Measurements of the sort
shown in Fig.~2(c) were repeated for the other fiber separations.
We then add all of these traces together to give the results shown
in Fig.~2(d). The two-photon interference fringes obtained by the
OCM method has about a 5.7-times larger fringe amplitude than the
QL results. The enhancement factor depends on the value of the
parameter $M$, which was 5.6 under our experimental conditions.
Using this $M$ value, the ratio between $C_{\rm PNR}$ and $C_{N
\rm PA}$ should be 3.3, but because the coincidence detection
efficiency of the QL detection system in Fig. 1(b) was reduced by
half due to the $1\times2$ coupler, the expected enhancement
factor becomes 5.6, in good agreement with the observed value. The
enhancement factor will increase if one uses a detector array of
smaller pixel size or if one enlarges the correlation area.

\begin{figure}
\centering
\includegraphics[width=8cm]{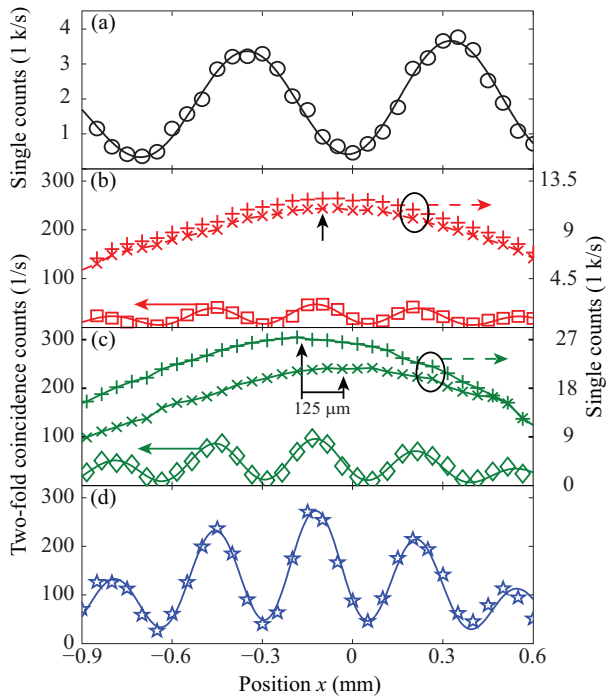}
\caption{(Color online) (a) Single-photon count rate for strongly
attenuated coherent-state light at an 800-nm wavelength versus the
detector position $x$.  (b)-(d) Single-photon (marked $+$ and
$\times$) and two-photon count rates for the $2002$ state as
measured by (b) the QL method of Fig.~1(b), (c) the OCM method of
Fig.~1(c), and (d) the OCM/PNR method described in the text.  For
part (c) the two parallel MMFs have a  separation of 125 $\mu$m.
The vertical arrows point the positions of the maximum
single-photon count rates. The solid lines are theoretical fits to
the data.  The integration time in each case was 10 seconds.}
\end{figure}

The OCM method is expected to scale well to higher values of $N$
and thus provide still greater spatial resolution.  However, the
implementation studied here based on the use of $N$ detectors of
variable separation provides a highly inefficient means of scaling
to higher $N$, because of the large number of detector
configurations that must be used.  Nonetheless, the results
presented  here provide a proof-of-principle demonstration that
the OCM method for $N=2$ can provide superresolution with a
two-fold enhancement over the classical resolution limit.  We have
also shown that the OCM method provides the same degree of
resolution enhancement as the QL method but with higher
efficiency.  We feel that, when large arrays of single-photon
detectors become available, the OCM method will be a powerful
means of providing still greater enhancement in resolution.

We thank Colin O'Sullivan for useful discussions.  This work was
supported by the U.S. Army Research Office through a MURI grant
and by the DARPA/DSO InPho program.

\bibliographystyle{unsrt}

\end{document}